\documentclass[aps,preprint,tightenlines]{revtex4}%
\usepackage{amsfonts}
\usepackage{amsmath}
\usepackage{amssymb}
\usepackage{graphicx}%
\providecommand{\U}[1]{\protect\rule{.1in}{.1in}}

\begin{document}
\title[Path integral approach to closed-form option pricing]{A path integral approach to closed-form option pricing formulas with
applications to stochastic volatility and interest rate models}
\author{D. Lemmens, M. Wouters, J. Tempere}
\altaffiliation{Also at: Lyman Laboratory of Physics, Harvard University, Cambridge MA 02138, USA.}

\affiliation{TFVS, Universiteit Antwerpen, Universiteitsplein 1, 2610 Antwerpen, Belgium.}
\author{S. Foulon}
\affiliation{KBC Bank, Havenlaan 12, 1080 Brussel, Belgium.}

\begin{abstract}
We present a path integral method to derive closed-form solutions for option
prices in a stochastic volatility model. The method is explained in detail for
the pricing of a plain vanilla option. The flexibility of our approach is
demonstrated by extending the realm of closed-form option price formulas to
the case where both the volatility and interest rates are stochastic. This
flexibility is promising for the treatment of exotic options. Our new
analytical formulas are tested with numerical Monte Carlo simulations.

\end{abstract}

\maketitle

\section{Introduction}

Since the seminal work of Black and Scholes \cite{BlackScholes,MERT}, who drew
an analogy between the random motion of microscopic particles and the
unpredictable evolution of stock prices, methods from theoretical physics have
proved very useful for pricing various financial derivative products
\cite{baaquiebook,kleinertbook,Dashb}. The pricing of derivative products is
based on a model for the evolution of the probability function of the
underlying asset. In order for a model to describe the economic reality
accurately, a sufficiently general evolution for the probability distribution
has to be allowed for. Nevertheless, the simple diffusion model of Black and
Scholes (BS) is still widely used. Much of its success is due to the
availability of closed-form analytical pricing formulas for many types of
derivatives \cite{wilmott}.

It is known for a long time that the BS model is only a crude approximation to
the economic reality and that its assumptions are violated in actual markets.
Perhaps the most illustrative violation is that the volatility implied from
traded vanilla options, the implied volatility is not constant across strikes
and maturities. Examples of models that tackle such violations are local
volatility processes \cite{Rcont,Derman}, jump processes \cite{Rcont},
L\'{e}vy processes \cite{Schoutens} and stochastic volatility models
\cite{Lipton}. A stochastic volatility model that has been particularly
successful at explaining the implied volatility smile in equity and foreign
exchange markets is the Heston model \cite{Heston}. In his seminal paper,
Heston \cite{Heston} derived a closed form solution for the price of a vanilla
option, which enables a quick and reliable calibration to market prices,
especially for liquidly traded vanilla options with maturities between 2
months and 2 years \cite{Cizek}. Contrary to the Black-Scholes model, to date
in the Heston model no closed-form analytic formulas have been found for
exotic options (for recent results see \cite{Griebsch}). Since no such
formulas are available in the literature for any but the simplest payoffs,
often costly numerical techniques have to be used (see \cite{Anders} and
references therein).

The original mathematical solution of the option pricing problem was
formulated within the framework of partial differential equations, but an
equivalent description with path integral methods was developed in the
pioneering work by Linetsky \cite{Linetsky} and Dash \cite{Dash,DashII}. They
showed that path dependent exotic options can be straightforwardly priced with
the path integral method. This should be intuitively clear: in the path
integral formalism, a probability is assigned to every evolution path of the
asset. In the formulation with partial differential equations, such quantities
are typically difficult to access.

Path integral methods have also been used in the pricing of options within
stochastic volatility models \cite{baaquieJP, Dragu} and in the related
problem of non-Gaussian diffusion \cite{KleinertPhysA} (at the end of Sec.
\ref{Pir} we come back to this connection), but to the best of our knowledge
no explicit option pricing formula as cheap to evaluate as Stein and Stein's
\cite{Stein} or Heston's formulas \cite{Heston} have yet been derived using
path integrals. We will show in the present paper how to carry out this task
for the Heston model. The result we thereby obtain corresponds to the existing
result \cite{Heston} for which the calibration and correspondence to market
data has already been investigated see for example
\cite{drag2,drag3,jap2,Yacine,Fior,Ranjan}. For a thorough discussion on when
which approach should be used we refer to \cite{Rebon} and references herein.

It is also known that there are still important features of asset price
distributions which are absent in the Heston model for example: empirical
studies of time series provide evidence of the long time memory of volatility
\cite{Bouch,Gop}. Since models containing a memory effect through retarded
interaction, for example in the context of polarons, \cite{FM}, have been
solved within a path integral framework, we think our method can prove to be
useful in more realistic models for the market also.

The full power of the path integral method becomes clear, when we exploit its
flexibility by calculating the price of an option in a setting where not only
the volatility but also the interest rate is stochastic and follows the widely
used CIR model \cite{cir,Brown,Gibbon,Dsev}. To the best of our knowledge, no
exact closed-form formula for this problem is available. Therefore, we have
checked our formulas against a Monte Carlo simulation.

The plan of the paper is as follows. In Sec. \ref{Pir}, we outline our model,
which is the one introduced by Heston. Extensions of our method to different
models are however straightforward. Further in this section we derive a
closed-form solution for the time evolution of the asset price. In Sec.
\ref{sec:PIvan} we present a closed-form pricing formula for plain vanilla
options which only involves one numerical integration of a compilation of
elementary functions. In Sec. III we will extend the Heston model to include
stochastic interest rate, in Sec. III A we present a closed-form solution for
the vanilla option price which still contains only on numerical integration of
a compilation of elementary functions. In Sec. \ref{Trosir}\ we test this
result with a Monte Carlo method and discuss the relevance of including
stochastic interest rate. Conclusions are drawn in Sec. \ref{conc}.

\section{Standard Heston model \label{sec:model}}

\subsection{The model and its path integral representation\label{Pir}}

We will concentrate on assets following a diffusion process described by the
following two equations introduced by Heston \cite{Heston}%

\begin{align}
dS  &  =\mu_{0}Sdt+S\sqrt{v}dw_{1},\label{Hest1a}\\
dv  &  =\kappa_{0}\left(  \theta_{0}-v\right)  dt+\sigma\sqrt{v}\left(  \rho
dw_{1}+\sqrt{1-\rho^{2}}dw_{2}\right)  . \label{Hest1}%
\end{align}
Here $S$ is the asset price, $\mu_{0}$ is a constant drift factor, $v$ is the
variance of the asset, $\kappa_{0}$ is the spring constant of the force that
attracts the variance to its mean reversion level $\theta_{0}$(also called the
mean reversion speed), $\sigma$\ is the volatility of the variance, and
$w_{1}$ and $w_{2}$ are independent Wiener processes with unit variance and
zero mean. The asset price follows a Black-Scholes process \cite{BlackScholes}%
, whereas the volatility obeys a Cox-Ingersoll-Ross process \cite{cir}.

There are two general approaches to determine the price of an option in a path
integral context. One could, based upon equations (\ref{Hest1a}),
(\ref{Hest1}) determine the probability distribution for the asset price at
the strike time $T$ conditional on the values of the asset and the variance at
the present time $P_{S}\left(  S_{T},v_{T}\mid S_{0},v_{0}\right)  $. The
expectation value of the option price at time $T$ can be calculated by
integrating the gain you make with a certain outcome of $S_{T}$ multiplied by
the probability of obtaining that outcome $P_{S}\left(  S_{T},v_{T}\mid
S_{0},v_{0}\right)  $\ over all possible values of $S_{T}$. To obtain the
present value of the price one then discounts this expectation value with the
risk free interest rate $r$. For a European call option this can be written
as:%
\begin{equation}
\mathcal{C}=e^{-rT}\int\limits_{-\infty}^{+\infty}dS_{T}dv_{T}\max\left[
S_{T}-K,0\right]  P_{S}\left(  S_{T},v_{T}\mid S_{0},v_{0}\right)  .
\label{jEcpf1}%
\end{equation}
We will refer to this approach as the "\emph{asset propagation approach}"
since $P_{S}$ is the propagator for a distribution of asset prices (and volatilities).

The other approach focuses on the option price rather than the asset
evolution, as will be referred to as the "\emph{option propagation approach}".
In his paper [6], Heston discusses the subtle differences between the asset
point of view and the option price point of view, and this discussion is also
relevant to the present path-integral framework. Heston motivates that the
time evolution of the option price $U\left(  S,v,t\right)  $ is governed by
the following partial differential equation (pde):%
\begin{equation}
\frac{\partial U}{\partial t}=-rS\frac{\partial U}{\partial S}+rU-\left\{
\kappa_{0}\left[  \theta_{0}-v\right]  -\lambda v\right\}  \frac{\partial
U}{\partial v}-\frac{1}{2}vS^{2}\frac{\partial^{2}U}{\partial S^{2}}%
-\rho\sigma vS\frac{\partial^{2}U}{\partial S\partial v}-\frac{1}{2}\sigma
^{2}v\frac{\partial^{2}U}{\partial v^{2}}, \label{Hestssde}%
\end{equation}
where $\lambda$ is a parameter introduced \cite{Heston} on the basis of
no-arbritage arguments and setting up a risk-free portfolio. If one makes the
substitution $U=e^{rt}V$ \ one obtains the following pde for $V$ as a function
of the asset price and the volatility:%
\begin{equation}
\frac{\partial V}{\partial t}=-rS\frac{\partial V}{\partial S}-\left\{
\kappa_{0}\left[  \theta_{0}-v\right]  -\lambda v\right\}  \frac{\partial
V}{\partial v}-\frac{1}{2}vS^{2}\frac{\partial^{2}V}{\partial S^{2}}%
-\rho\sigma vS\frac{\partial^{2}V}{\partial S\partial v}-\frac{1}{2}\sigma
^{2}v\frac{\partial^{2}V}{\partial v^{2}}. \label{Hestssde2}%
\end{equation}
Based on this pde, one can find a kernel $P_{V}$ \ that propagates a given
final distribution $V\left(  S_{T},v,T\right)  $ backwards to the present
value $V(S_{0},v_{0},0)$ of the option. Since the value of the option at the
final time $T$ is known, $V\left(  S_{T},v,T\right)  =e^{-rT}U\left(
S_{T},v,T\right)  =e^{-rT}\max\left[  S_{T}-K,0\right]  $, the value of the
option now is obtained through
\begin{equation}
\mathcal{C}=e^{-rT}\int\limits_{-\infty}^{+\infty}dS_{T}dv_{T}\max\left[
S_{T}-K,0\right]  P_{V}\left(  S_{T},v_{T}\mid S_{0},v_{0}\right)  ,
\label{Ecpf2}%
\end{equation}
Furthermore the pde (\ref{Hestssde2}) is equal to the Kolmogorov backward
equation corresponding to the following system of stochastic differential
equations
\begin{align}
dS  &  =rSdt+S\sqrt{v}dw_{1},\\
dv  &  =\left\{  \kappa_{0}\left[  \theta_{0}-v\right]  -\lambda v\right\}
dt+\sigma\sqrt{v}\left(  \rho dw_{1}+\sqrt{1-\rho^{2}}dw_{2}\right)  .
\end{align}

This means that both approaches can be dealt with simultaneously by
considering a generalized stochastic process:%
\begin{align}
dS  &  =\mu Sdt+S\sqrt{v}dw_{1},\label{gensde1}\\
dv  &  =\kappa\left(  \theta-v\right)  dt+\sigma\sqrt{v}\left(  \rho
dw_{1}+\sqrt{1-\rho^{2}}dw_{2}\right)  . \label{gensde2}%
\end{align}
and calculating its transition probability $P\left(  S_{T},v_{T}\mid
S_{0},v_{0}\right)  $. The "asset propagation" approach (\ref{jEcpf1}) can
then be retained by simply replacing $\mu,\kappa$ and $\theta$\ by $\mu
_{0},\kappa_{0}$ and $\theta_{0}$ and the "option propagation" approach
(\ref{Hestssde2}), (\ref{Ecpf2}) by replacing $\mu,\kappa$ and $\theta$\ by
$r,\kappa_{0}+\lambda$ and $\kappa_{0}\theta_{0}/\left(  \kappa_{0}%
+\lambda\right)  $. The pricing formula for the European call is the same as
(\ref{jEcpf1})\ where this time the transition probability $P\left(
S_{T},v_{T}\mid S_{0},v_{0}\right)  $ is the one corresponding to
(\ref{gensde1}), (\ref{gensde2}):%
\begin{equation}
\left\{
\begin{array}
[c]{l}%
P_{S}\left(  S_{T},v_{T}\mid S_{0},v_{0}\right)  =\left.  P\left(  S_{T}%
,v_{T}\mid S_{0},v_{0}\right)  \right\vert _{\mu=\mu_{0},\kappa=\kappa
_{0},\theta=\theta_{0}}\\
P_{V}\left(  S_{T},v_{T}\mid S_{0},v_{0}\right)  =\left.  P\left(  S_{T}%
,v_{T}\mid S_{0},v_{0}\right)  \right\vert _{\mu=r,\kappa=\kappa_{0}%
+\lambda,\theta=\kappa_{0}\theta_{0}/(\kappa_{0}+\lambda)}%
\end{array}
\right.  \label{vbern}%
\end{equation}
We will calculate the transition density $P$ for the general stochastic
process (\ref{gensde1}), (\ref{gensde2}).

For later convenience we make the following substitutions:%
\begin{align}
x  &  =\ln\left(  \frac{S}{S_{0}}\right)  -\mu t,\label{xdefb}\\
z  &  =\sqrt{v},\nonumber
\end{align}
$x$ is called the logreturn and $z$ is the volatility of the asset price.
After these substitutions, Eq. (\ref{Hest1})\ becomes:%
\begin{align}
dx  &  =-\frac{z^{2}}{2}dt+zdw_{1},\label{Hest3a}\\
dz  &  =\left[  \frac{1}{2z}\left(  \kappa\theta-\frac{\sigma^{2}}{4}\right)
-\kappa\frac{z}{2}\right]  dt+\frac{\sigma}{2}\left(  \rho dw_{1}+\sqrt
{1-\rho^{2}}dw_{2}\right)  . \label{Hest3}%
\end{align}
The substitution
\[
y\left(  t\right)  =x\left(  t\right)  -\frac{\rho}{\sigma}\left(
z^{2}\left(  t\right)  -\kappa\theta t\right)  ,
\]
leads to two uncorrelated equations:%
\begin{align}
dy  &  =\left(  \frac{\rho}{\sigma}\kappa-\frac{1}{2}\right)  z^{2}%
dt+z\sqrt{1-\rho^{2}}dw_{1},\label{Hest4a}\\
dz  &  =\left[  \frac{1}{2z}\left(  \kappa\theta-\frac{\sigma^{2}}{4}\right)
-\kappa\frac{z}{2}\right]  dt+\frac{\sigma}{2}dw_{2}. \label{Hest4}%
\end{align}
We will assume that the initial volatility $z(t=0)=z_{0}$ is known
\cite{Cizek}. The probability that $y$\ has the value $y_{T}$ and $z$ the
value $z_{T}$\ at a later time $T$ will be denoted as $P\left(  y_{T}%
,z_{T}\mid y_{0},z_{0}\right)  $. The advantage of transforming to these
variables is that $dw_{1}$ and $dw_{2}$ are uncorrelated, so that the
following expression holds for $P\left(  y_{T},z_{T}\mid y_{0},z_{0}\right)
$:%
\begin{equation}
P\left(  y_{T},z_{T}\mid y_{0},z_{0}\right)  =\int\mathcal{D}y\mathcal{D}%
z\exp\left(  -\int\limits_{0}^{T}\left\{  \mathcal{L}_{Q}\left[
y(t),z(t)\right]  +\mathcal{L}_{CIR}[z(t)]\right\}  dt\right)  .
\label{PImooi}%
\end{equation}
Where the quadratic Lagrangian $\mathcal{L}_{Q}\left(  y(t),z(t)\right)
$\ equals
\begin{equation}
\mathcal{L}_{Q}\left(  y(t),z(t)\right)  =\frac{1}{2z^{2}\left(  1-\rho
^{2}\right)  }\left[  \dot{y}-\left(  \frac{\rho}{\sigma}\kappa-\frac{1}%
{2}\right)  z^{2}\right]  ^{2}, \label{elkuu}%
\end{equation}
and the Lagrangian corresponding to the CIR process, $\mathcal{L}_{CIR}%
[z(t)]$,\ is given by \cite{Rosa}:%
\begin{equation}
\mathcal{L}_{CIR}[z]=\frac{2}{\sigma^{2}}\left\{  \dot{z}-\frac{1}{2}\left[
\frac{1}{z}\left(  \kappa\theta-\frac{\sigma^{2}}{4}\right)  -\kappa z\right]
\right\}  ^{2}-\frac{1}{4z^{2}}\left(  \kappa\theta-\frac{\sigma^{2}}%
{4}\right)  -\frac{\kappa}{4}, \label{cird}%
\end{equation}

The first step in the evaluation of Eq. (\ref{PImooi}) is the integration over
all $y$-paths. Because the action is quadratic in $y$ this integration can be
done analytically and yields
\begin{align}
P\left(  y_{T},z_{T}\mid y_{0},z_{0}\right)   &  =\int\mathcal{D}z(t)\frac
{1}{\sqrt{2\pi\bar{z}^{2}\left(  1-\rho^{2}\right)  }}\exp\left\{
\frac{\left(  \frac{\rho}{\sigma}\kappa-\frac{1}{2}\right)  }{\left(
1-\rho^{2}\right)  }\left(  y_{T}-y_{0}\right)  \right. \nonumber\\
&  -\frac{1}{2\left(  1-\rho^{2}\right)  }\left(  \frac{\rho}{\sigma}%
\kappa-\frac{1}{2}\right)  ^{2}\bar{z}^{2}-\frac{\left(  y_{T}-y_{0}\right)
^{2}}{2\bar{z}^{2}\left(  1-\rho^{2}\right)  }\nonumber\\
&  \left.  -\int_{0}^{T}dt\text{ }\mathcal{L}_{CIR}[z(t)]\right\}  .
\label{eq:PI1b}%
\end{align}
Note that the probability to arrive in $(y_{T},z_{T})$ only depends on the
average value of the volatility along the path $z(t)$: $\bar{z}^{2}=\int
_{0}^{T}z^{2}(t)dt$, in agreement with Ref. \cite{Stein}. However, this
average value appears in the denominator of the third term, and to perform the
functional integral one needs to bring this into the numerator. This is
achieved by rewriting part of the expression (\ref{eq:PI1b}) as follows:%
\begin{equation}
\frac{1}{\sqrt{2\pi\bar{z}^{2}\left(  1-\rho^{2}\right)  }}\exp\left[
-\frac{\left(  y_{T}-y_{0}\right)  ^{2}}{2\bar{z}^{2}\left(  1-\rho
^{2}\right)  }\right]  =\int_{-\infty}^{+\infty}\frac{dk}{2\pi}\exp\left[
i\left(  y_{T}-y_{0}\right)  k-\frac{\int z^{2}dt\left(  1-\rho^{2}\right)
}{2}k^{2}\right]  . \label{eq:gidb}%
\end{equation}
Combining Eqns. (\ref{eq:PI1b}) and (\ref{eq:gidb}) and making the
substitution $k=l+i\frac{\left(  \frac{\rho}{\sigma}\kappa-\frac{1}{2}\right)
}{\left(  1-\rho^{2}\right)  }$ the transition probability becomes
\begin{align}
P\left(  y_{T},z_{T}\mid y_{0},z_{0}\right)   &  =\int_{-\infty}^{+\infty
}\frac{dl}{2\pi}\exp\left[  i\left(  y_{T}-y_{0}\right)  l\right]
\int\mathcal{D}z(t)\label{eq:PI2b}\\
\times &  \exp\left(  -\int_{0}^{T}dt\left\{  \mathcal{L}_{CIR}[z(t)]+\frac
{1}{2}z^{2}\left[  \left(  1-\rho^{2}\right)  l^{2}+2li\left(  \frac{\rho
}{\sigma}\kappa-\frac{1}{2}\right)  \right]  \right\}  \right)  .\nonumber
\end{align}
The path integral over the CIR action is formally equivalent to the exactly
solvable radial harmonic oscillator \cite{GenS} and, fortunately, adding terms
proportional to $z^{2}$ to the action does not spoil this equivalence. The
full path integral over $z(t)$ can be carried out without approximations with
the following result:%
\begin{align}
P\left(  y_{T},z_{T}\mid y_{0},z_{0}\right)   &  =\frac{1}{2\pi}\exp\left[
\frac{\kappa^{2}\theta}{\sigma^{2}}T+\left(  2\frac{\kappa\theta}{\sigma^{2}%
}-\frac{1}{2}\right)  \ln\left(  \frac{z_{T}}{z_{0}}\right)  -\frac{\kappa
}{\sigma^{2}}\left(  z_{T}^{2}-z_{0}^{2}\right)  \right] \nonumber\\
\times &  \int\limits_{-\infty}^{+\infty}\exp\left[  i\left(  y_{T}%
-y_{0}\right)  l\right]  \sqrt{z_{0}z_{T}}\frac{4\omega}{\sigma^{2}%
\sinh\left(  \omega T\right)  }\nonumber\\
\times &  \exp\left[  -\frac{2\omega}{\sigma^{2}}\left(  z_{0}^{2}+z_{T}%
^{2}\right)  \coth\left(  \omega T\right)  \right]  I_{\frac{2}{\sigma^{2}%
}\kappa\theta-1}\left[  \frac{4\omega z_{0}z_{T}}{\sigma^{2}\sinh\left(
\omega T\right)  }\right]  dl. \label{napad2}%
\end{align}
where
\begin{equation}
\omega=\frac{\sigma}{2}\sqrt{\left(  \frac{\kappa}{\sigma}+il\rho\right)
^{2}+l\left(  l-i\right)  }. \label{denomega}%
\end{equation}
is the $l$-dependent frequency of the radial harmonic oscillator that
corresponds to the CIR Lagrangian (\ref{cird}). After transforming back to the
$x$ variable we see that also the integral over the final value $z_{T}$ can be
done analytically (see e.g. \cite{Prudni}), yielding the marginal probability
distribution $\mathcal{P}\left(  x_{T}\mid0,z_{0}\right)  =\int_{-\infty
}^{+\infty}dz_{T}P\left(  x_{T},z_{T}\mid0,z_{0}\right)  $ (written in the
original variable $v$) as a simple Fourier integral:%
\begin{align}
\mathcal{P}\left(  x_{T}\mid0,v_{0}\right)   &  =\frac{1}{2\pi}\exp\left[
\frac{\kappa}{\sigma^{2}}\left(  \kappa\theta T+v_{0}\right)  \right]
\nonumber\\
&  \times\int\limits_{-\infty}^{+\infty}N^{\frac{2}{\sigma^{2}}\kappa\theta
}\exp\left\{  i\left[  x_{T}+\frac{\rho}{\sigma}\left(  v_{0}+\kappa\theta
T\right)  \right]  l\right. \nonumber\\
&  \left.  -\frac{2\omega}{\sigma^{2}\sinh\left(  \omega T\right)  }\left[
\cosh\left(  \omega T\right)  -N\right]  v_{0}\right\}  dl, \label{eq:Pfinc}%
\end{align}
where N is:
\begin{equation}
N=\frac{1}{\cosh\left(  \omega T\right)  +\frac{1}{2\omega}\left(
\kappa+il\rho\sigma\right)  \sinh\left(  \omega T\right)  }. \label{eq:Pfinb}%
\end{equation}
Note the similarity of the expression (\ref{eq:Pfinc}) with the result
obtained in Ref. \cite{KleinertPhysA}, derived for a general stochastic
process with non-Gaussian noise.

\subsection{Pricing of plain vanilla options \label{sec:PIvan}}

From now on we follow the option propagation approach and set $\mu$ equal to
$r$. The price of a call option $\mathcal{C}$ with expiration date $T$ and
strike $K$ when the transition probability is known is given by Eq.
(\ref{jEcpf1}). Writing this formula in the $x$ variable and thereby inserting
the result (\ref{eq:Pfinc}) for the transition probability results in:%
\begin{equation}
\mathcal{C}=e^{-rT}\int\limits_{-\infty}^{+\infty}dx_{T}\max\left[  S_{0}%
\exp\left(  x_{T}\right)  -K,0\right]  \mathcal{P}\left(  x_{T}\mid
0,v_{0}\right)  , \label{eq:priceP}%
\end{equation}
where the risk free interest rate was restored and denoted by $r$. Now there
are still two numerical integrations that have to be done. Following the
derivation outlined in Ref. \cite{KleinertPhysA}\ we can rewrite expression
(\ref{eq:priceP}) so that only one numerical integration remains:%
\begin{align}
\mathcal{C}  &  =\frac{S_{0}-e^{-rT}K}{2}+i\int\limits_{-\infty}^{\infty}%
\frac{1}{l}\left\{  \exp\left[  i\left(  \frac{\rho}{\sigma}a+x_{e}-rT\right)
l+\frac{\kappa}{\sigma^{2}}a\right]  \right. \label{resvop}\\
&  \left.  \times\left[  S_{0}\exp\left(  \Theta-\frac{\rho}{\sigma}a\right)
-e^{-rT}K\exp\left(  \Upsilon\right)  \right]  -S_{0}+e^{-rT}K\right\}
\frac{dl}{2\pi},\nonumber
\end{align}
with
\begin{subequations}
\begin{align}
x_{e}  &  =\ln\left(  \frac{K}{S_{0}}\right)  ,\label{eerst}\\
a  &  =v_{0}+\kappa\theta T,\\
\nu &  =\frac{\sigma}{2}\sqrt{\left(  \frac{\kappa}{\sigma}+il\rho
-\rho\right)  ^{2}+l\left(  l+i\right)  },\\
M  &  =\left[  \cosh\left(  \nu T\right)  +\frac{1}{2\nu}\left(  \kappa
+il\rho\sigma-\rho\sigma\right)  \sinh\left(  \nu T\right)  \right]  ^{-1},\\
\Theta &  =\frac{2\nu v_{0}}{\sigma^{2}\sinh\left(  \nu T\right)  }\left[
M-\cosh\left(  \nu T\right)  \right]  +\frac{2}{\sigma^{2}}\kappa\theta\log
M,\\
\Upsilon &  =\frac{2\omega v_{0}}{\sigma^{2}\sinh\left(  \omega T\right)
}\left[  N-\cosh\left(  \omega T\right)  \right]  +\frac{2}{\sigma^{2}}%
\kappa\theta\log N. \label{laatst}%
\end{align}
and $\omega$ defined as before (\ref{denomega}). We\ have tested this result
against the formula stated in Ref. \cite{Heston}. This confirmed the
correctness of formula (\ref{resvop}).\ Now we are confident to explore new
grounds with our method in the following section.

\section{Stochastic interest rate}

\subsection{Derivation of the option price}

In the previous section we assumed the interest rate to be constant. Here we
allow the interest rate to change in time, $r(t)$. Applying Black and Scholes'
no-arbitrage argument on Heston's risk-free portfolio motivation for the
evolution of the option price, we again obtain the partial differential
equation (\ref{Hestssde2}) with $r(t)$ rather than a constant $r:$
\end{subequations}
\begin{equation}
\frac{\partial V}{\partial t}=-r(t)S\frac{\partial V}{\partial S}-\left\{
\kappa_{0}\left[  \theta_{0}-v\right]  -\lambda v\right\}  \frac{\partial
V}{\partial v}-\frac{1}{2}vS^{2}\frac{\partial^{2}V}{\partial S^{2}}%
-\rho\sigma vS\frac{\partial^{2}V}{\partial S\partial v}-\frac{1}{2}\sigma
^{2}v\frac{\partial^{2}V}{\partial v^{2}}%
\end{equation}
For a given function $r(t)$ this leads to a kernel $P_{V}\left[  S_{T}%
,v_{T}\mid S_{0},v_{0}\mid r(t)\right]  $ so that the option price becomes%
\begin{equation}
\mathcal{C}[r(t)]=\int\limits_{-\infty}^{+\infty}dS_{T}dv_{T}\max\left[
S_{T}-K,0\right]  \text{ }e^{-\int r(t)dt}P_{V}\left[  S_{T},v_{T}\mid
S_{0},v_{0}\mid r(t)\right]  .
\end{equation}
Note that the option price is now a functional of the given time evolution of
the interest rate $r(t)$. As in the previous section, it is convenient to
introduce new integration variables
\begin{align}
y\left(  t\right)   &  =\ln\left(  \frac{S}{S_{0}}\right)  -\frac{\rho}%
{\sigma}\left[  z^{2}\left(  t\right)  -\kappa\theta t\right]  ,\\
z(t)  &  =\sqrt{v(t)}.
\end{align}
In the path-integral treatment, the kernel can be written as a sum over all
possible realizations of $y(t)$ and $z(t)$, weighed by the action functional
of the system:
\begin{align}
\mathcal{C}[r(t)]  &  =\int\limits_{-\infty}^{+\infty}dx_{T}dv_{T}\max\left[
e^{x_{T}}-K,0\right]  \text{ }e^{-%
{\textstyle\int\nolimits_{0}^{T}}
r(t)dt}\nonumber\\
&  \times\int\mathcal{D}y\mathcal{D}z\text{ }\exp\left(  -\int\limits_{0}%
^{T}\left\{  \mathcal{L}_{Q}\left[  y(t),z(t),r(t)\right]  +\mathcal{L}%
_{CIR}[z(t)]\right\}  dt\right)  ,
\end{align}
where $\mathcal{L}_{Q}$ is the quadratic Lagrangian (\ref{elkuu})
\begin{equation}
\mathcal{L}_{Q}\left(  y(t),z(t)\right)  =\frac{1}{2z^{2}\left(  1-\rho
^{2}\right)  }\left[  \dot{y}(t)-r(t)-\left(  \frac{\rho}{\sigma}\kappa
-\frac{1}{2}\right)  z^{2}(t)\right]  ^{2},
\end{equation}
and $\mathcal{L}_{CIR}$ is the CIR\ Lagrangian. Of course, we cannot know what
particular realization of the interest rate $r(t)$ will appear in the future.
We assume the interest rate to follow a CIR process which is uncorrelated from
the other two stochastic processes,%
\begin{equation}
dr=\kappa_{r}\left(  \theta_{r}-r\right)  dt+\sigma_{r}\sqrt{r}dw_{3}.
\end{equation}
The value for the option price then needs to be averaged over the realization
of $r(t)$ in this CIR\ process. Where the calculation of the expectation value
of such a functional might become cumbersome with conventional probabilistic
techniques, it can be evaluated very elegantly with the Feynman-Kac formula:
\begin{equation}
\mathcal{C}=\left\langle \mathcal{C}[r(t)]\right\rangle =\int\mathcal{D}%
r\text{ }\mathcal{C}[r(t)]\exp\left(  -\int\limits_{0}^{T}\mathcal{L}%
_{CIR}[r(t)]dt\right)  ,
\end{equation}
where $\mathcal{L}_{CIR}$ is the Lagrangian for the CIR process. The final
result can be expressed with a modified propagator $P(S_{T},v_{T},r_{T}\mid
S_{0},v_{0},r_{0})$ as%
\begin{equation}
\mathcal{C}=\int\limits_{-\infty}^{+\infty}dS_{T}dv_{T}dr_{T}\max\left[
S_{T}-K,0\right]  P(S_{T},v_{T},r_{T}\mid S_{0},v_{0},r_{0}), \label{blab}%
\end{equation}
with%
\begin{align}
&  P(S_{T},v_{T},r_{T}|S_{0},v_{0},r_{0})=\int\mathcal{D}y\mathcal{D}%
z\mathcal{D}r\text{ }e^{-\int_{0}^{T}r\left(  t\right)  dt}\nonumber\\
&  \times\exp\left(  -\int\limits_{0}^{T}\left\{  \mathcal{L}_{Q}\left[
y(t),z(t),r(t)\right]  +\mathcal{L}_{CIR}[z(t)]+\mathcal{L}_{CIR}%
[r(t)]\right\}  dt\right)  .
\end{align}
The stochastic interest rate makes the vanilla price dependent on the specific
path followed by the interest rate. This part of the payoff has been taken
into the calculation of the propagator, where it is analytically tractable,
and no longer appears explicitly in the expression (\ref{blab}) for the option
price. Herein lies the strength of the path-integral approach, to price
path-dependent options. With a stochastic interest rate the European vanilla
option becomes dependent on the entire path of the interest rate and is still
solved in a very straightforward way. This is promising for more general
option types, such as the barrier and Asian options that we are currently investigating.

A useful substitution to perform the functional integrations is%
\begin{align}
\vartheta_{1}(t)  &  =\sqrt{r(t)},\nonumber\\
\vartheta_{2}(t)  &  =y\left(  t\right)  -\int_{0}^{t}r(t^{\prime})dt^{\prime
}. \label{subsngv}%
\end{align}
As was the case for the Lagrangian corresponding to the volatility, the
Lagrangian corresponding to the interest rate process will also be formally
equivalent to the Lagrangian corresponding to a radial harmonic oscillator;
furthermore the addition of another term quadratic in $\vartheta_{1}$ stemming
from the discount factor doesn't spoil the correspondence. The result reads as
follows:%
\begin{align}
\mathcal{C}  &  =\frac{1}{2}\left[  S_{0}-K\exp\left(  \frac{\kappa_{r}%
}{\sigma_{r}^{2}}a_{r}+\Upsilon_{r}\left(  0\right)  \right)  \right]
\nonumber\\
&  +i\int\limits_{-\infty}^{\infty}\frac{1}{l}\left\{  K\exp\left[
\Upsilon_{r}\left(  0\right)  +\frac{\kappa_{r}}{\sigma_{r}^{2}}a_{r}\right]
\right.  -S_{0}+\exp\left[  i\left(  \frac{\rho}{\sigma}a+x_{e}\right)
l+\frac{\kappa}{\sigma^{2}}a+\frac{\kappa_{r}}{\sigma_{r}^{2}}a_{r}\right]
\nonumber\\
&  \left.  \times\left[  S_{0}\exp\left(  -\frac{\rho}{\sigma}a+\Theta
+\Theta_{r}\right)  -K\exp\left(  \Upsilon+\Upsilon_{r}\right)  \right]
\right\}  \frac{dl}{2\pi}. \label{prmr2}%
\end{align}
To make it surveyable, we introduced the following notations%
\begin{align*}
a_{r}  &  =r_{0}+\kappa_{r}\theta_{r}T,\\
\nu_{r}  &  =\frac{\sigma_{r}}{2}\sqrt{\frac{\kappa_{r}^{2}}{\sigma_{r}^{2}%
}+2il},\\
\omega_{r}\left(  l\right)   &  =\frac{\sigma_{r}}{2}\sqrt{\frac{\kappa
_{r}^{2}}{\sigma_{r}^{2}}+2\left(  il+1\right)  },\\
M_{r}  &  =\left[  \cosh\left(  \nu_{r}T\right)  +\frac{\kappa_{r}}{2\nu_{r}%
}\sinh\left(  \nu_{r}T\right)  \right]  ^{-1},\\
\Theta_{r}  &  =\frac{2\nu_{r}r_{0}}{\sigma_{r}^{2}\sinh\left(  \nu
_{r}T\right)  }\left[  M_{r}-\cosh\left(  \nu_{r}T\right)  \right]
+2\frac{\kappa_{r}\theta_{r}}{\sigma_{r}^{2}}\log M_{r},\\
\Upsilon_{r}\left(  l\right)   &  =\frac{2\omega_{r}\left(  l\right)  r_{0}%
}{\sigma_{r}^{2}\sinh\left[  \omega_{r}\left(  l\right)  T\right]  }\left\{
N_{r}\left(  l\right)  -\cosh\left[  \omega_{r}\left(  l\right)  T\right]
\right\}  +2\frac{\kappa_{r}\theta_{r}}{\sigma_{r}^{2}}\log N_{r}\left(
l\right)  .
\end{align*}
These notations reflect the extension to the case of stochastic interest rate
(symbols with subscript $r$) of the corresponding quantities in the Heston
model (equations (\ref{eerst})-(\ref{laatst})). Notice the resemblance with
formula (\ref{resvop}). Formula (\ref{prmr2}) still contains\ just one
numerical integration with an integrand composed out of elementary functions.
To the best of our knowledge, only approximate analytical formulae are
available when both the volatility and interest rate are stochastic
\cite{Jap}. Because of the lack of alternative exact analytical expressions,
we have checked the validity of our formula (\ref{prmr2}) against numerical
Monte Carlo simulations. Our Monte Carlo method is outlined below.

First notice that substitutions (\ref{subsngv}) transform the $x$-variable
into a variable $\tilde{x}$, independent of the interest rate by subtracting
the time averaged interest rate $\bar{r}$: $\tilde{x}=x-\bar{r}$. This results
in the same equation as in the constant interest rate situation, Eq.
(\ref{Hest3a}). Also the discount factor only contains $\bar{r}$. This means
that the knowledge of the probability distribution $\bar{r}$ is sufficient to
calculate the price by means of the formula (\ref{resvop}) derived in the
constant interest rate setting. So the Monte Carlo scheme used is the
following: first values for $\bar{r}$ are simulated and used to calculate the
option price for these values, next the price is averaged over all the
simulations. A value for $\bar{r}$ is simulated as follows: time is
discretized in little time steps $\Delta$, we sample a path for $r$ and
integrate along this path. To calculate the probability distribution for
$\bar{r}$, we used the result that the stochastic time increment of a CIR
variable over a small time step $\Delta t$ follows a non-central $\chi^{2}$
distribution \cite{Anders}. The probability distribution of the average
interest rate $\bar{r}$ is then simulated by generating many $r$-paths in
discretized time. As shown in Fig. \ref{prentsti5}, the agreement between the
analytical (thick full line) and numerical option prices is excellent.

In this section the option propagation approach was followed from the
beginning. In this setting it is necessary to make a choice between the two
approaches from the start because in the asset propagation approach one would
actually have to introduce a stochastic process for the drift $\mu_{0}%
$\ instead of for the interest rate. That these two should follow the same
stochastic process is not clear. Since the option propagation approach is the
most common one anyway we followed this approach. If one does want to
introduce a stochastic process for the drift $\mu_{0}$\ this poses no problem
and the derivation of an option price in this setting would be completely similar.

\subsection{Results and discussion.\label{Trosir}}

In the current treatment, we have two layers of generalization as compared to
the Black-Scholes result. First, the volatility appearing in the Black-Scholes
process is stochastic -- this leads to the Heston model. Second, the interest
rate of the Black-Scholes model is also stochastic -- leading to our present
results. In this paragraph, we argue that both improvements can have an
equally important effect on the option price.%
\begin{figure}
[t]
\begin{center}
\includegraphics[
height=1.8559in,
width=6.0744in
]%
{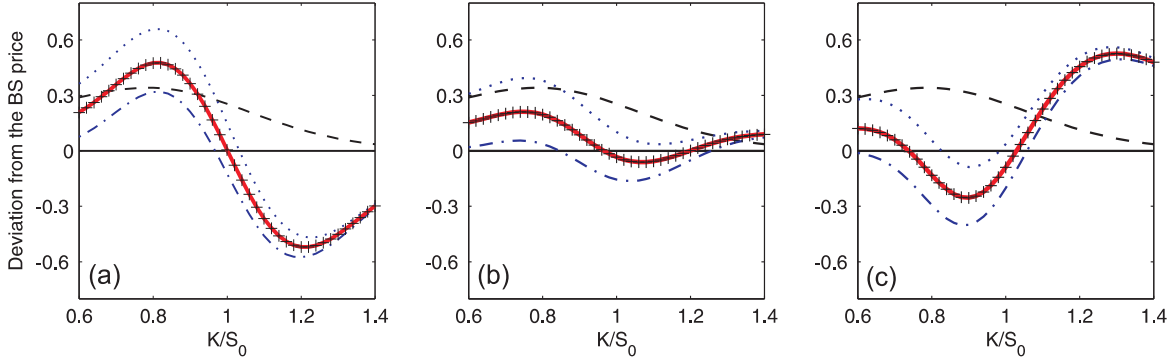}%
\caption{(color online, two column wide) This figure shows the result of
different pricing formulas from which the Black-Scholes result (with interest
rate $r=\theta_{r}$) has been subtracted. Since we are not considering a
specific asset, the option price could be stated in any currency, therefore
the deviation is given in arbitrary units. The thick (red) curve shows our
analytical results for the model with both stochastic interest rate $r$ and
stochastic volatility. The crosses represent results from a Monte Carlo
simulation of our model, confirming the analytical formula. The blue dotted
curve and the blue dash-dotted curve show the results for the Heston model
with constant interest rate $r=r(0)$ and $r=\theta_{r}$, respectively. The
dashed curve shows the results for a Black-Scholes model with $r=r(0).$ The
following parameter values were used for all three panels: $\kappa=1$,
$\sigma=0.2$, $\theta=0.04$, $v_{0}=0.04$, $T=1$, $S_{0}=100$, $\kappa
_{r}=1.8$; $\sigma_{r}=0.1$; $\theta_{r}=0.03$; $r_{0}=0.035$. The correlation
coefficient is for panel (a) $\rho=-0.5,$ (b) $\rho=0$ and (c) $\rho=0.5$. }%
\label{prentsti5}%
\end{center}
\end{figure}

This is illustrated in Fig. \ref{prentsti5}, where the different approaches
are compared. Let's start with the most complete model, where both interest
rate and volatility are stochastic. The resulting option price, Eq.
(\ref{prmr2}), for this model is shown as a thick red curve. The result from
the closed-form expression agrees well with the Monte Carlo simulation, shown
as crosses.

\begin{figure}
[ptb]
\begin{center}
\includegraphics[
height=1.9156in,
width=2.9603in
]%
{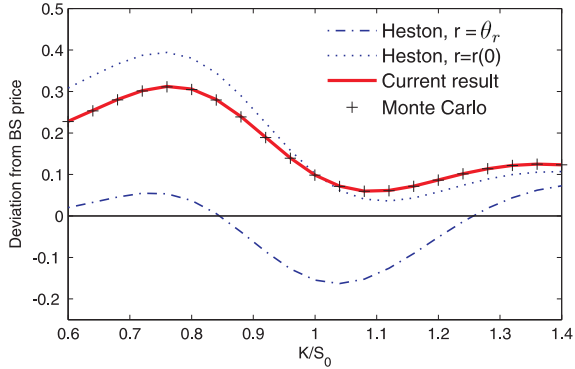}%
\caption{(color online). As in Fig. 1, the result of different pricing
formulas from which the Black-Scholes result (with interest rate $r=\theta
_{r}$, thin black line) has been subtracted, is shown. The following parameter
values were used: $\kappa=1$, $\sigma=0.2$, $\theta=0.04$, $v_{0}=0.04$,
$T=1$, $S_{0}=100$, $\kappa_{r}=0.5$, $\sigma_{r}=0.3$, $\theta_{r}=0.03$,
$r_{0}=0.035$, $\rho=0$. }%
\label{figje2}%
\end{center}
\end{figure}

Now we strip off one layer of complexity, and fix the interest rate $r$ -- it
is no longer a stochastic variable. Then we obtain the Heston model as
an\ `approximation' to a stochastic interest rate world. The question poses
itself of which fixed interest rate to use, if we want to make the comparison.
Two choices are shown in Fig. \ref{prentsti5}: $r=r(0)$ and $r=\theta_{r}$.
The former choice (dotted blue curves) sets the Heston interest rate equal to
the interest rate at time $0$, whereas the latter choice (dash-dotted curves)
sets the Heston interest rate equal to the mean reversion level $\theta_{r}$.
For the parameter values used in Fig. \ref{prentsti5}, the most complete
result lies between the two Heston `approximations', but this is not
necessarily so. Fig. \ref{figje2} shows that for some choices of other
(realistic) parameters, the full result can lie outside both Heston
approximations. Nevertheless, as $\kappa$ becomes very large, the stochastic
interest rate will be drawn very tightly to the mean reversion rate
$\theta_{r}$, and one expects the full result to be near the Heston
approximation with $r=\theta_{r}$. When $\kappa$ is very small, the stochastic
interest rate will not be drawn quickly towards $\theta_{r}$ so that when also
$\sigma_{r}$ is small, the full results will be near the Heston approximation
with $r=r(0)$.

Next, we strip off the second layer of approximation, and also fix the
volatility. This results in the familiar Black-Scholes model as the crudest
approximation to our system. Now a second choice has to be made: which value
of the volatility to use. Here, we take the stochastic volatility at time zero
to be equal to the mean reversion level of the volatility CIR\ process, so
that the ambiguity of choice is avoided. The choice for what interest rate to
use, however, remains. In Fig. \ref{prentsti5}, we show the Black-Scholes
results with $r=r(0)$ (dashed line) and $r=\theta_{r}$ (full line). We have
plotted all the results relative to the Black-Scholes result with
$r=\theta_{r}$ to emphasize the differences rather than the absolute magnitude
of the prices (for this reason, the $r=\theta_{r}$ Black-Scholes result is the
baseline of the plots). The difference between the three panels of Fig.
\ref{prentsti5} is the value of the correlation between asset price and
volatility.%

From Figs. \ref{prentsti5} and \ref{figje2}, it is clear that both levels of
approximation (keeping the volatility constant and keeping the interest rate
constant) have an equally large effect on the option price. Even within the
Heston framework, the choice of what value to use for the interest rate is
seen to influence the price considerably. Choosing a different interest rate,
or keeping the interest rate as a stochastic variable, leads to a price
correction that is as large as the price correction obtained by going from the
Black-Scholes to the Heston model. This result emphasizes the importance of a
correct treatment of the interest rate in pricing models (This also depends
strongly on the length of the lifetime of the option).

Finally we must remark that the price differences when working within the
standard Heston model or within the extended one can be influenced by the
calibration method. For Figures. \ref{prentsti5} and \ref{figje2} we used the
same parameters for the volatility process both in the standard model and in
the extended one, parameter values for the interest rate process are
calibrated separately. Literature shows that the parameter values for the
volatility process (see for example \cite{jap2} and \cite{Heston}) and the
interest rate process (see for example \cite{Dsev} and \cite{Brown}) can
attain values in a broad range containing the values we chose to produce Fig.
\ref{prentsti5} and Fig. \ref{figje2}. However if the parameter values
obtained for the interest rate process are used in formula (\ref{prmr2}) to
calibrate the remaining parameter values for the volatility process one might
get different results. We can not exclude that this calibration approach would
lead to smaller price differences between the two approaches. However such a
calibration is a research area on its own and is outside the scope of this article.

\section{Conclusions \label{conc}}

We have developed a path-integral method to derive closed-form analytical
formulas for the asset price distribution in the Heston stochastic volatility
model. Closed-form formulas are obtained for the logreturn of the derivative
and the vanilla option price. The presented results correspond to the known
semi-analytic results obtained from solving the partial differential equation
\cite{Heston} by standard techniques.

The flexibility of our approach is demonstrated by extending the results to
the case where the interest rate is a stochastic variable as well, and follows
a CIR process. For this case, to the best of our knowledge, no exact
analytical solutions have been derived before. We have checked our
semi-analytical results for the model with both stochastic volatility and
stochastic interest rate against a Monte-Carlo simulation. The quantitative
analysis shows that the effect of stochastic interest rate on the Heston model
can be as large as the effect of the stochastic volatility on the
Black-Scholes model. However we did not perform a full calibration, which
might influence the results. Finally, the analogy between stochastic interest
rate models and path dependent options makes our method promising for the
pricing of exotic derivative products.

\begin{acknowledgments}
Acknowledgments -- Discussions with L. Lemmens, I. De Saedeleer, K. in't Hout
and E. Boksenbojm are gratefully acknowledged. This work is supported
financially by the Fund for Scientific Research - Flanders, FWO project
G.0125.08. J. T. and D. L. gratefully acknowledge support of the Special
Research Fund of the University of Antwerp, BOF NOI UA 2007.
\end{acknowledgments}


\begin{thebibliography}{99}                                                                                               %


\bibitem {BlackScholes}F. Black and M. Scholes, J. Pol. Econ. \textbf{81} 637 (1973).

\bibitem {MERT}R. C. Merton, Bell Journal of Economics and Management Science,
\textbf{4}, 141 (1973)

\bibitem {baaquiebook}B. E. Baaquie, \emph{Quantum Finance: Path Integrals and
Hamiltonians for Options and Interest Rates }(Cambridge University Press,
Cambridge, 2004).

\bibitem {kleinertbook}H. Kleinert, \emph{Path Integrals in Quantum Mechanics,
Statistics, Polymer Physics, and Financial Markets}, (World Scientific,
Singapore, 2004).

\bibitem {Dashb}J. Dash, Quantitative Finance and Risk Management: A
Physicist's Approach.

\bibitem {wilmott}P. Wilmott, J. Dewynne, and S. Howison, \emph{Option
Pricing} (Oxford Financial Press, Oxford, 1993).

\bibitem {Rcont}R. Cont, P. Tankov, Financial modelling with jump processes,
Chapman \& Hall/CRC, (2003).

\bibitem {Derman}E. Derman, I. Kani, Riding on a smile, Risk 7 (1994) 32-39.

\bibitem {Schoutens}W. Schoutens, Levy Processes in Finance: Pricing Financial
derivatives, Wiley, 2003.

\bibitem {Lipton}A. Lipton, Mathematical Methods for Foreign Exchange: A
Financial Engineer's Approach, World Scientific, 2001.

\bibitem {Heston}S. L. Heston,\textsl{ }Review of Financial Studies
\textbf{6}, 327 (1993).

\bibitem {Cizek}P. Cizek, W. H\"{a}rdle and R. Weron, Statistical Tools for
Finance and Insurance, Springer, 2005.

\bibitem {Griebsch}S. Griebsch, Pricing of Exotic options in Heston's
stochastic volatility model, Lecture, Frankfurt Mathfinance Workshop, March 2007.

\bibitem {Anders}L. B. G. Andersen, "Efficient Simulation of the Heston
Stochastic Volatility Model" (January 23, 2007). Available at SSRN: http://ssrn.com/abstract=946405.

\bibitem {Linetsky}V. Linetsky, Computational Economics \textbf{11}, 129 (1998).

\bibitem {Dash}J. Dash, \textsl{Path integrals and options: Part I.} (CNRS
preprint CPT-88/PE.2206, 1988).

\bibitem {DashII}J. Dash, \textsl{Path integrals and options: Part II.} (CNRS
preprint CPT-89/PE.2333, 1989).

\bibitem {Dragu}A. A. Dr\u{a}gulescu, arXiv:cond-mat/0307341.

\bibitem {baaquieJP}B. E. Baaquie, J. Phys. I France \textbf{7}, 1733 (1997).

\bibitem {KleinertPhysA}H. Kleinert, Physica A \textbf{338}, 151 (2004).

\bibitem {Stein}E. M. Stein and J. C. Stein, Review of Financial Studies
\textbf{4}, 727 (1991).

\bibitem {drag2}A. A. Dr\u{a}gulescu and V. M. Yakovenko Quantitative Finance
\textbf{2} 443-453 (2002).

\bibitem {drag3}A. C. Silva and V. M. Yakovenko Physica A \textbf{324} 303 --
310 (2003).

\bibitem {jap2}J. E. Zhang and J. Shu, Proceedings of the IEEE International
Conference on Computational Intelligence for Financial Engineering, 2003,
pages 85-92.

\bibitem {Yacine}Y. A\"{\i}t-Sahalia and R. Kimmel Journal of Financial
Economics volume 83 issue 2 pages 413-452 (2007).

\bibitem {Fior}G. Fiorentini, A. Le\'{o}n and G. Rubio Journal of Empirical
Finance volume 9 issue 2 pages 225-255 (2002).

\bibitem {Ranjan}S. R. Das and R. K. Sundaram, Journal of Financial and
Quantitative Analysis volume 34 issue 2 pages 211-239 (1999).

\bibitem {Rebon}R. Rebonato, Volatility and Correlation: the Perfect Hedger
and the Fox, 2e Edition (John Wiley, Chichester, 2004).

\bibitem {Bouch}J. P. Bouchaud and M. Potters \emph{Theory of Financial Risk
and Derivative Pricing }(Cambridge University Press, New York, 2003).

\bibitem {Gop}P. Gopikrishnan, V. Plerou, L. A. N. Amaral, M. Meyer and H. E.
Stanley PRE 60, 5 (1999).

\bibitem {FM}R. P. Feynman ; Statistical Mechanics: A Set of Lectures,
Advanced Book Classics, 2e Edition (Perseus Books Group 1998).

\bibitem {cir}J.C. Cox , J.E. Ingersoll, and S.A. Ross, Econometrica
\textbf{53}, 385 (1985).

\bibitem {Brown}R. H. Brown and S. M. Schaefer, Journal of Financial
Economics, Volume 35, Issue 1, pages 3-42 (1994).

\bibitem {Gibbon}M. R. Gibbons and K. Ramaswamy, The Review of Financial
Studies, Volume 6, Issue 3 pages 619-658 (1993).

\bibitem {Dsev}D. \v{S}ev\v{c}ovi\v{c} and A. Urb\'{a}nov\'{a} Csajkov\'{a},
Cejor \textbf{13} 169-188 (2005).

\bibitem {Rosa}E. Bennati, M. Rosa-Clot, and S. Taddei, International Journal
of Theoretical \& Applied Finance \textbf{2}, 381 (1999); also at arXiv:cond-mat/9901277.

\bibitem {GenS}C. Grosche and F.Steiner, \textsl{Handbook of Feynman path
integrals.}\ (Springer, Berlin, 1998).

\bibitem {Prudni}A. P. Prudnikov, Y. A. Brychkov, and O. I. Marichev
\textsl{Integrals and series, volume 2: special functions }(Gordon and Breach,
New York, 1992).

\bibitem {Jap}N. Kunitomo and Y. J. Kim, The Japanese Economic Review
\textbf{58}, 71 (2007).
\end{thebibliography}
\end{document}